\documentclass{CUP-JNL-DCE}

\usepackage{latexsym}
\usepackage{graphicx}
\usepackage{multicol,multirow}
\usepackage{amsmath,amssymb,amsfonts}
\usepackage{mathrsfs}
\usepackage{amsthm}
\usepackage{apacite}
\usepackage{rotating}
\usepackage{appendix}
\usepackage[authoryear]{natbib}
\usepackage{ifpdf}
\usepackage[T1]{fontenc}
\usepackage{times}
\usepackage{sourcesanspro}
\usepackage{newtxmath}
\usepackage{textcomp}%
\usepackage{xcolor}%
\usepackage{hyperref}
\usepackage[nameinlink,capitalize]{cleveref}
\usepackage[normalem]{ulem}
\graphicspath{{figures/}}
\usepackage{amsmath,amsfonts,amssymb}
\usepackage{bm}
\DeclareMathOperator*{\argmax}{arg\,max}
\usepackage[inline]{enumitem}
\usepackage{subcaption}
\usepackage{multicol}
\usepackage{xcolor}
\usepackage[normalem]{ulem}
\usepackage[nameinlink,capitalize]{cleveref}

\articletype{POSITION PAPER}
\jname{Data-Centric Engineering}
\jyear{2024}

\begin{document}

\begin{Frontmatter}

\title{Monitoring-Supported Value Generation for Managing Structures and Infrastructure Systems}

\author*[1]{Antonios Kamariotis}\email{antonios.kamariotis@ibk.baug.ethz.ch}
\author[1]{Eleni Chatzi}
\author[2]{Daniel Straub}
\author[3]{Nikolaos Dervilis}
\author[4]{Kai Goebel}
\author[3]{Aidan J. Hughes}
\author[5]{Geert Lombaert}
\author[6]{Costas Papadimitriou}
\author[7]{Konstantinos G. Papakonstantinou}
\author[8]{Matteo Pozzi}
\author[9]{Michael Todd}
\author[3]{Keith Worden}

\authormark{A. Kamariotis \textit{et al.}}

\address[1]{\orgdiv{Institute of Structural Engineering}, \orgname{ETH Zurich}, \orgaddress{\street{Stefano-Franscini-Platz 5}, \postcode{8093 Zurich}, \country{Switzerland}}}

\address[2]{\orgdiv{Engineering Risk Analysis Group \& Munich Data Science Institute}, \orgname{Technical University of Munich}, \orgaddress{\street{Arcisstr. 21}, \postcode{80333 Munich}, \country{Germany}}}

\address[3]{\orgdiv{Dynamics Research Group, Department of Mechanical Engineering}, \orgname{University of Sheffield},\orgaddress{\street{} \postcode{Sheffield, S1 3JD}, \country{UK}}}

\address[4]{\orgdiv{SRI International/PARC},\orgname{} \orgaddress{\street{Palo Alto,} \postcode{CA 94304}, \country{USA}}}

\address[5]{\orgdiv{Department of Civil Engineering}, \orgname{KU Leuven}, \orgaddress{\street{Leuven}, \postcode{}\country{Belgium}}}

\address[6]{\orgdiv{Department of Mechanical Engineering}, \orgname{University of Thessaly}, \orgaddress{\street{Pedion Areos}, \postcode{38334}, \country{Greece}}}

\address[7]{\orgdiv{Department of Civil \& Environmental Engineering}, \orgname{The Pennsylvania State University}, \orgaddress{\street{University Park}, \postcode{PA}, \country{USA}}}

\address[8]{\orgdiv{Civil and Environmental Engineering}, \orgname{Carnegie Mellon University}, \orgaddress{\street{Pittsburgh}, \postcode{PA}, \country{USA}}}

\address[9]{\orgdiv{}\orgname{University of California San Diego}, \orgaddress{\street{9500 Gilman Drive}, \postcode{La Jolla, CA 92093}, \country{USA}}}


\abstract{To maximize its value, the design, development and implementation of Structural Health Monitoring (SHM) should focus on its role in facilitating decision support. In this position paper, we offer perspectives on the synergy between SHM and decision-making. We propose a classification of SHM use cases aligning with various dimensions that are closely linked to the respective decision contexts. The types of decisions that have to be supported by the SHM system within these settings are discussed along with the corresponding challenges. We provide an overview of different classes of models that are required for integrating SHM in the decision-making process to support management and operation and maintenance of structures and infrastructure systems. Fundamental decision-theoretic principles and state-of-the-art methods for optimizing maintenance and operational decision-making under uncertainty are briefly discussed. Finally, we offer a viewpoint on the appropriate course of action for quantifying, validating and maximizing the added value generated by SHM. This work aspires to synthesize the different perspectives of the SHM, Prognostic Health Management (PHM), and reliability communities, and deliver a roadmap towards monitoring-based decision support.}

\begin{policy}[Impact Statement]
Structural Health Monitoring (SHM) systems can be viewed as decision-support tools. This position paper aims to deliver a roadmap towards monitoring-supported value creation and systematic integration of SHM in the operation and maintenance decision-making process for structures and infrastructure systems.
\end{policy}

\end{Frontmatter}

\section{Introduction}
\label{sec:introduction}

Structural Health Monitoring (SHM) offers a potent tool to enhance the Operation \& Maintenance (O\&M) decision-making process for structures and infrastructure systems \citep{Farrar_Worden_2013}. SHM systems can essentially be viewed as a collection of tools for decision support \citep{HUGHES2021107339, KAMARIOTIS2023109708}. Yet, to date, SHM systems have not been broadly deployed on real-world structures and infrastructure systems \citep{Cawley_2018}. This is not least because of the fact that the decision-support potential of SHM remains relatively unexplored. SHM research has been mainly driven by technological and methodological advancements without explicitly taking into consideration insights and methods from the risk/reliability and decision-making communities. Few recent efforts have been made to connect these two lines of research and formally explore the Value of Information (VoI) stemming from SHM \citep{Pozzi_2011, Thons_2018, HUGHES2021107339, KAMARIOTIS2023109708}. In the experience of the authors, different research fields comprise different perspectives, often entailing distinct vocabularies, which renders exchange challenging. Moreover, owners and operators of structures and infrastructure systems and SHM practitioners must be convinced in order to adopt advanced SHM technologies and trust decision algorithms for assisting them in the O\&M process. A paradigm shift is required, as typically the O\&M process heavily depends on a rule-based philosophy and is strongly regularized.  Actionable use cases are required for illustrating the manner in which SHM systems can support different decision settings, thereby generating a return on investment.

In this position paper, experts from the SHM, risk/reliability and decision-making, as well as the Prognostic Health Management (PHM) communities, jointly offer perspectives on the synergistic  development of SHM and decision-making tools, and discuss multiple associated challenges. Specifically, in \cref{sec:SHM_uses_cases} we propose a classification of SHM use cases across different dimensions that feed the O\&M decision-making process. \cref{sec:models} overviews both purely data-driven and hybrid diagnostic/prognostic models and discusses how these facilitate a monitoring-informed maintenance planning process. \cref{sec:decision_making} discusses the optimization of maintenance planning strategies under availability of monitoring data, and offers an overview of the Value of Information (VoI) framework. In \cref{sec:quantifying}, we discuss approaches and directions towards increased value generation with SHM. Finally, \cref{sec:concluding} offers brief concluding remarks.

\section{SHM use cases in relation to the decision-making process}
\label{sec:SHM_uses_cases}

\begin{figure}
\centering
\includegraphics[width=0.98\textwidth]{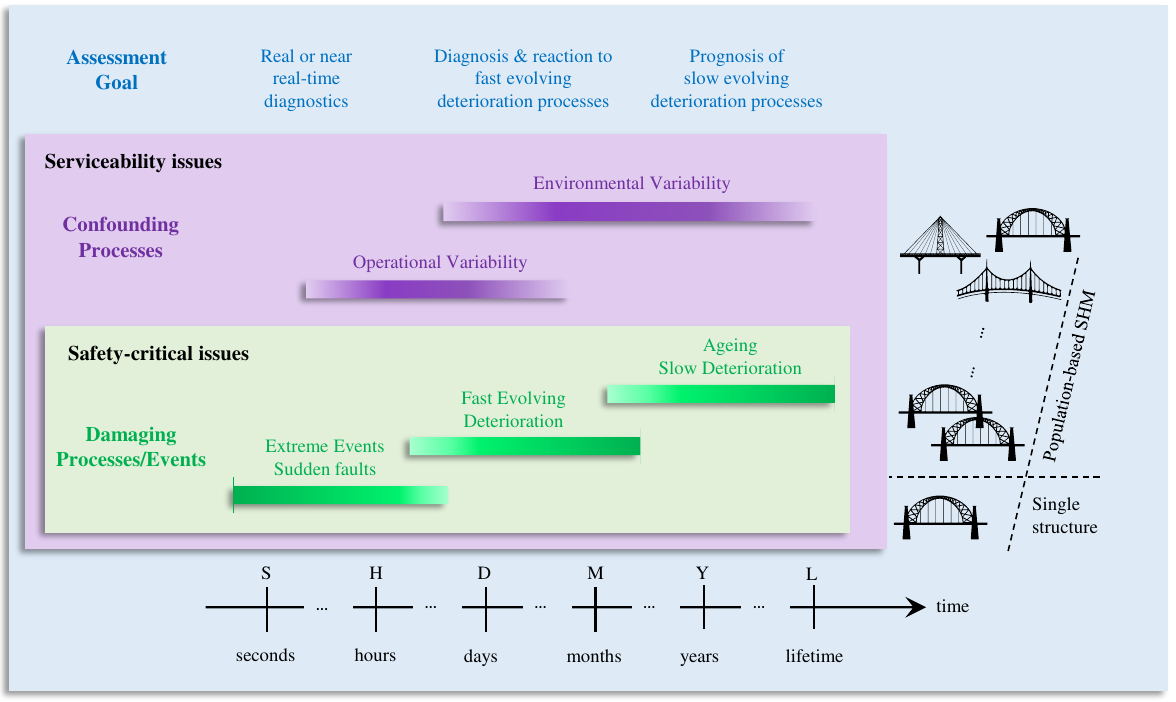}
\caption{SHM use cases across dimensions that influence decision-making for monitored structures}
\label{fig:SHM_dimensions}
\end{figure}

SHM finds application in various use cases, each associated with different contexts in which decisions can be supported by the monitoring data and SHM processing algorithms. We here discuss some main dimensions along which SHM use cases can be classified, with each dimension describing different aspects that influence the O\&M decision-making task. These dimensions are described in \cref{subsec:time_scales,subsec:structural_performance,subsec:PBSHM} and are illustrated in \cref{fig:SHM_dimensions}. For each dimension, we further discuss specific decision settings and associated challenges.
\subsection{Monitoring across temporal scales}\label{subsec:time_scales}
\cite{KAMARIOTIS2023109708} present a classification of SHM use cases in terms of the associated time scales for decision-making, ranging from real-time needs (sub-second accuracy) to decisions spanning the lifetime of the system. At the lower end of the scale (seconds to hours), SHM can be deployed to deliver real-time or near real-time diagnostics. The assessment goal at this time scale is the fast, almost online, detection of flaws or abnormalities. Examples include the real-time detection of a sudden fault in a control system, or the flagging of structures after an earthquake event \citep{Tubaldi_2022}. At the scale of days to months, an objective of SHM is to identify fast-evolving structural deterioration processes that could compromise serviceability or safety, such as, e.g., freeze-thaw or akali-silica reaction processes. Furthermore, at this scale, SHM can serve for post-event assessment following extreme events (e.g., floods) or devising appropriate remedial strategies. Finally, at the scale of larger time spans (years to lifetime), SHM can be leveraged to support condition-based or predictive maintenance decisions associated with slow-evolving deterioration processes. Assessment is effectuated via use of both purely data-driven or hybrid schemes, where either data exclusively or data coupled with physics-based models are used to form twinning, diagnostic and prognostic  tools (see \cref{sec:models}). Such assessment is typically impaired by presence of confounding processes, typically reflected in Environmental and Operational Variability (EOV) (see \cref{fig:SHM_dimensions}) \citep{Peeters_2001,Figueiredo_2011,Cross_2011}. The time scale determines the urgency of the decision-making task, which consequently dictates the type of models and decision-making strategies that can be employed, as discussed in detail in \cref{sec:models,sec:decision_making} respectively. 

Currently, SHM systems are most typically called upon to assist in closer examination of existing systems that exhibit identified potential problems, or to support decisions related to lifetime extension, with sensors deployed at an advanced stage of the structural life-cycle. However, SHM can also be integrated during the design phase, escorting structural systems from cradle to grave \citep{Farrar_Worden_2013,Hulse_2020}. Depending on the stage in which installation or extension of a monitoring deployment is contemplated, Value of Information (VoI) or Value of SHM (VoSHM) analyses (see \cref{sec:decision_making}) can inform the decision on whether or not to install a specific SHM system on a target structure, as well as advise on the configuration of the monitoring deployment for maximizing the associated VoI, according to the objective at hand.
\subsection{Monitoring for varying structural performance requirements}
\label{subsec:structural_performance}

Opportunities and challenges for SHM-supported decision making depend on the criticality of the monitored processes. In structural engineering, it is common to distinguish between serviceability requirements, which relate to ensuring the intended use of the structure, and safety requirements, which relate to ensuring the safe operation of the system. The latter are typically much stricter and require compliance with regulations and codes, which are often rule-based. It can be difficult for owners and operators to deviate from these rules, and in these cases SHM cannot reduce the cost of prescribed inspections and maintenance, unless it is also taken into account in forming new guidelines. In some application areas, e.g., in earthquake engineering, performance-based requirements are increasingly considered, but these are still the exception rather than the norm. 
By contrast, regulations for serviceability requirements are typically less strict and owners of structures have some latitude on how to to ensure serviceability. Hence it can be easier currently to include SHM into the decision process when dealing with servicability issues. 
This differentiation between serviceability and safety issues can also affect the type and urgency of associated maintenance actions. 

\subsection{SHM for individual structures versus population-based SHM}
\label{subsec:PBSHM}

SHM use cases depend also on the scale of the monitored object(s) and the associated decision making, which can be at the component, individual asset, and eventually the population (fleet) level. A population-based approach to SHM has recently been introduced in a series of contributions \citep{BULL2021107141,GOSLIGA2021107144,GARDNER2021107142,TSIALIAMANIS2021107692}. Population-based SHM (PBSHM) is characterised by the sharing of information between sufficiently-similar structures, with the aim of improving predictive performance and decision-making. PBSHM can mitigate the problem of data scarcity for individual structures, which prevents full exploitation of supervised learning in data-driven approaches. Population-based approaches to SHM extend the use cases of monitoring systems to supporting O\&M decision-making for {\em fleets} of structures -- this extension is reflected as a third dimension in \cref{fig:SHM_dimensions}.
Another consideration for population-based SHM is that the deployment of full-scale monitoring systems for all structures within a population may be too costly, or that diminishing returns are seen in terms of VoSHM as more members of the population are subject to full-scale monitoring. Therefore, it may be preferable to target a few salient structures with full-scale SHM systems, and rely on reduced-scale monitoring in conjunction with transfer learning \citep{5288526} (or other information-sharing technologies) in order to support decision-making for the remaining structures in the population. 
Finally, it is worth noting that the value of PBSHM includes a component associated with the expected utility gained as a result of sharing, or transferring, information between structures. This quantity, termed the (expected) value of information \emph{transfer}, is useful to consider as it can be used to select optimal algorithms and parameters to conduct transfer learning \citep{Hughes2024quantifying}.
\subsection{Industry and regional culture}

The industrial culture is an important aspect to consider. In many industries, a rule-based methodology is typically followed in the practical O\&M process. For instance, for bridge structures, inspections and maintenance actions are typically scheduled based on a fixed period (e.g., inspections every 2-3 years), as required by codes and standards \citep{Bridge_inspectors_manual}. Adoption of SHM for enhancing the O\&M process requires transition to a performance-based methodology; this  is followed to a certain extent in O\&M of wind turbines \citep{General_Electric}. Furthermore, there exist different degrees of regularization in terms of maintenance requirements and practices across different countries and regions, which has an impact on the feasibility of a paradigm shift regarding the O\&M process.
\section{Model classes for SHM-based assessment}
\label{sec:models}

\begin{figure}
	\centering	\includegraphics[width=0.75\textwidth]{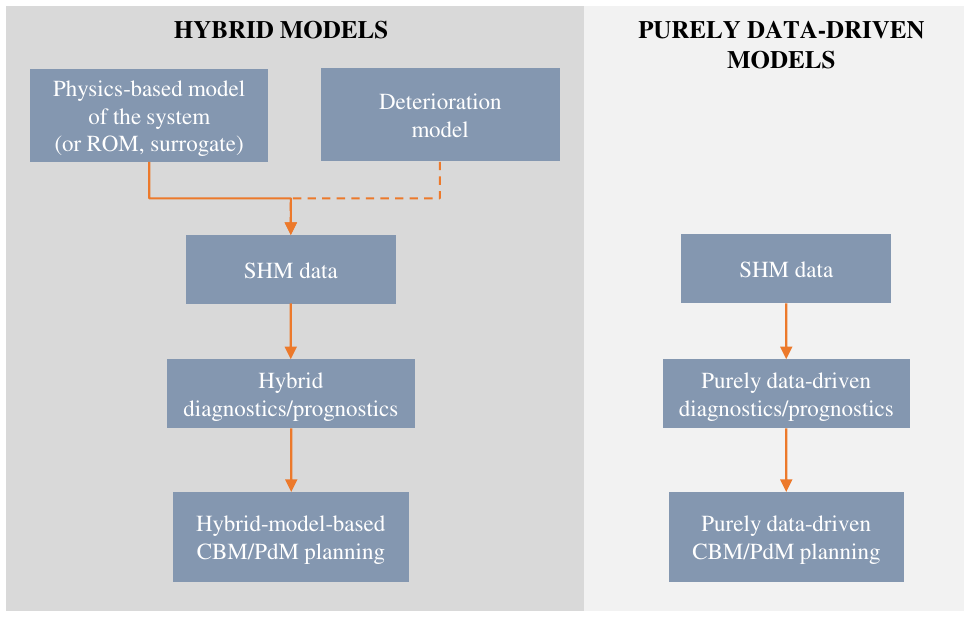}
	\caption{Modeling layers required for SHM-aided operation and maintenance planning}
	\label{fig:Model layers}
\end{figure}

This section describes the different model classes that are required for integrating SHM in the O\&M decision-making process, as illustrated in \cref{fig:Model layers}. 
One can distinguish between at least two distinct maintenance planning paradigms that are enabled by SHM: i) condition-based maintenance (CBM) and ii) predictive maintenance (PdM) planning \citep{Fink2020, Goebel_2017}. Following a CBM strategy, maintenance is informed at the moment when a threshold is exceeded, imposed either directly on the value of an observation or on the value of a damage indicator, which relates to the current condition of the system. Instead, a PdM strategy relies on prognostic models, developed also through SHM and related data, which deliver predictions of the future evolution of a system's condition, with maintenance informed on the basis of these predictions. While a CBM-based approach is accomplished on the basis of availability of data and the frequent accompanying use of related models, a PdM-based approach often requires the inclusion of a physics- or engineering-based model in the loop. This holds particularly in instances where the systems being monitored lack sufficient experimental data to failure. While such data may be available for certain standardized (non-unique) engineering systems (e.g., in industrial engineering) \citep{ZIO2022108119}, they are typically not available for safety-critical civil and infrastructure engineering systems, which generally feature more individual designs. 

When no models are available \textit{a-priori}, SHM data alone can be utilized for the training of \textit{purely data-driven models}, also referred to as \textit{black-box models}, for damage diagnosis/prognosis. Purely data-driven models can be inferred via the use of system identification \citep{SID} and/or machine learning (ML) schemes \citep{Farrar_Worden_2013,Malekloo_2022}. 
Data-driven models cannot easily move away from existing experience and thus typically fail to extrapolate to future predictions regarding the evolution of damage, i.e., they fail to extend from damage diagnosis to prognosis \citep{Farrar_Worden_2013}. The effective development of purely data-driven prognostic models relies on datasets that contain monitoring data corresponding to damage states of several systems similar to the system of interest \citep{NASA}, which poses a challenge for structures, where designs are individualized. PBSHM attempts to tackle this challenge by capitalizing on partial similarities of such structures for generalizing models and transferring the knowledge gained from monitoring several such individual instances \citep{tsialiamanis_2023}. To date, the purely data-driven PdM planning paradigm has enjoyed broader application within the Prognostic Health Management (PHM) discipline \citep{NGUYEN2019251,Fink2020,LEE2023108908,KAMARIOTIS2024109723}, with its application to structures and infrastructure systems remaining scarce.

When physics-based models are available (e.g., through finite element (FE) models), these can offer valuable intuition into the underlying system. In this case, SHM data can be fused with physics-based models, resulting in \textit{hybrid} or \textit{grey-box} representations, which can serve diagnostic, prognostic, and eventually, decision support tasks \citep{ARIASCHAO2022107961,LIU2022109276,cross2023spectrum}. A hybrid model can refer to i) estimators of quantities of interest that incorporate physics principles (e.g., physics-informed Gaussian Processes \citep{cross2023spectrum}) or ii) full-blown \textit{Digital Twins} (DTs) \citep{Wagg_2020,Chinesta2020,VANDERHORN2021113524,Thelen2022}. When a DT is constructed offline, via a one-off model updating process \citep{SIMOEN2015123}, then this can be viewed as a \textit{Digital Twin Instance} (DTI) \citep{McClellan_2022}. When such an updating process is executed continually, this can be viewed as a closed loop DT, also referred to as DT aggregate. It reflects an aggregation of DTIs, which allows tracking the structure (physical twin) throughout its life-cycle and informing decisions that realize value \citep{AIAA}. In the particular case where data and models can be fused on the fly, as data are attained, we refer to a \textit{Real-Time Digital Twin} (RTDT) \citep{VETTORI2023109654}. Referring to the prior categorization of SHM use cases across time scales, it becomes evident that detailed engineering models can typically not be harnessed for real-time or near real-time tasks. In tackling this challenge, Reduced Order Models (ROMs) \citep{Benner_2015,Chinesta_2016,VLACHAS2021116055}, or surrogate representations \citep{Luthen_2021}, form invaluable tools that reduce the computational complexity of detailed engineering models and allow for real-time tasks to be accomplished. 

When considering the environment/context within which a system operates, a hybrid setting requires prior knowledge on the types of damage and the mathematical definition of deterioration processes acting on the monitored structure or infrastructure. This knowledge is typically embedded in the \textit{a-priori} definition of empirical or physics-based deterioration models \citep{VANNOORTWIJK20092,JIA201899}. However, adequate quantitative deterioration models exist only for a small subset of deterioration phenomena acting on structures and infrastructures \citep{Straub_2018}. This issue poses significant challenges, as such models are indispensable for making predictions. A strong need therefore exists for the development of improved deterioration models, a process that can be assisted by the existence of monitoring data. If such models are available, they can be updated on-line, or even partially inferred, based on the monitoring data, thus delivering data-informed predictions of the damage evolution \citep{Straub_2009,ZIO2011403,CORBETTA2018305,kamariotis_sardi_papaioannou_chatzi_straub_2023,MORATO2023109144}. This assimilation of deterioration models and SHM data essentially also forms part of hybrid modeling. In this context, a PdM strategy can be employed based on input from hybrid prognostics, which rely on availability of a physics-based model of the engineering system and a deterioration model, and their fusion with SHM data. 
\section{Decision-making under uncertainty with SHM}
\label{sec:decision_making}

SHM-aided maintenance planning is a problem of decision-making under uncertainty. It can be performed following a CBM or a PdM decision strategy. A decision strategy $S$ consists of a set of decision rules adopted at all time steps of a sequential decision problem, specifying the action(s) to take (among a predefined set of actions) at each time step.

A decision strategy may be assigned \textit{a-priori}, i.e., a decision maker may suboptimally assign a threshold on the value of a damage diagnostic/prognostic indicator, which, when exceeded, will inform a maintenance action. Fixing a decision strategy \textit{a-priori} requires prior engineering expertise regarding the considered system. This approach best reflects what is typically done in practice, where a formal strategy optimization is rarely performed.

In the case of systems for which monitoring data from damage states from similar systems exist, the \textit{a-priori} definition of a decision strategy is certainly assisted by the existence of such data. Furthermore, when prior information is available, and the systems are standardized (e.g., wind turbines, rotating machinery, gearboxes, bearings) the \textit{a-priori} definition of a decision strategy is somewhat easier. On a side note, a benefit of SHM adoption is that it can also assist with general knowledge generation in this process, for both standardized and non-standardized systems.
\subsection{Optimizing a decision strategy}

A decision strategy may be acquired through an optimization process. Optimizing a decision strategy under uncertainty is performed by means of the \textit{principle of maximum expected utility} \citep{Berger_1985}. The goal is to identify the optimal decision strategy $S^{*}$ that maximizes the expected utility:
\begin{equation}
	S^{*} = \argmax_{S\in\mathscr{S}}\mathbb{E}_{\boldsymbol{X}}[U(S,\boldsymbol{X})],
\label{eq:expected_utility}
\end{equation}
where $\mathbb{E}_{\boldsymbol{X}}$ is the mathematical expectation with respect to the uncertain model parameters $\boldsymbol{X}$ and $U(S,\boldsymbol{X})$ is the utility over the decision time horizon when strategy $S$ is implemented. The definition of a decision time horizon is problem-dependent. In the context of structure and infrastructure engineering systems, the time horizon of interest is often the total life-cycle. In some cases, the total life-cycle can also be decided based on the decision strategy, i.e., decommission/termination action or replacement.
\subsubsection{Bayesian decision analysis}

In Bayesian decision theory, the uncertain state of the environment is characterized by the vector $\boldsymbol{X}$, which includes uncertain quantities of the physics-based model of the system, uncertain parameters of the deterioration model, describing the temporal evolution of damage, and/or the uncertain damage state. It is assumed that a prior probabilistic model of $\boldsymbol{X}$ is available to the analyst. When the expectation in the strategy optimization of \cref{eq:expected_utility} is performed with respect to the prior distribution of $\boldsymbol{X}$, one refers to a \textit{prior decision analysis}, whereby the SHM data are not accounted for. As discussed above, when SHM data are merged with physics-based models of the system and/or deterioration models, we refer to hybrid models. In such a hybrid setting, SHM data can be used to update the knowledge and reduce the uncertainty about the probabilistic model of $\boldsymbol{X}$. This updating is performed via Bayesian inference \citep{Gilks_1995,särkkä_svensson_2023}. The definition of a likelihood function is required for Bayesian inference; it relates the condition of the system to the data obtained with the monitoring system \citep{bismut2022unifying}. Examples of likelihood functions are, e.g, a probability of detection curve for damage detection \citep{Long_2022}, or a probabilistic model of the discrepancy between identified and model-predicted eigenfrequencies \citep{BEHMANESH2015360}. Once the probabilistic model of $\boldsymbol{X}$ is updated based on SHM data, the expectation in the strategy optimization of \cref{eq:expected_utility} can be performed with respect to the posterior distribution of $\boldsymbol{X}$. This renders a \textit{posterior decision analysis}, such as in the example of miter gate structures that comprise inland waterway corridors \citep{Vega22}, and in the example of railway infrastructure \citep{ARCIERI2023109496}. Recent work has also extended modeling human risk perception in the decision process that biases information provided by SHM \citep{Chadha23}.
\subsubsection{Value of Information (VoI) analysis}
\label{subsec:VOI}

SHM data become available only after the actual installation and operation of a SHM system. Nonetheless, one is often interested in investigating and quantifying the potential economic benefit associated with SHM adoption in an operational evaluation level. Such investigations can be performed within the framework of a Value of Information (VoI) analysis \citep{Pozzi_2011,Straub_2014,Thons_2018,Nielsen_2021,Kamariotis_2022,Giordano2022,ZHANG2022102258}, which entails a preposterior Bayesian decision analysis \citep{Raiffa_1961}. A VoI analysis quantifies the expected improvement in decision-making due to the reduced uncertainty offered by information sources. Specifically, the VoI metric is quantified by the difference in expected total utility with and without the SHM system. More specialized metrics, such as the Value of SHM (VoSHM) \citep{Andriotis_2021,KAMARIOTIS2023109708} and the normalized expected reward-to-risk ratio \citep{Chadha22} have been introduced.
A VoI/VoSHM analysis relies on simulation of future scenarios, hence requiring a dedicated probabilistic model of the investigated SHM system. This model is required for generating monitoring data that one expects to extract from this SHM system for multiple sampled trajectories (sampled from $\boldsymbol{X}$). Such a data generation process may be facilitated by use of a probabilistic digital twin \citep{tsialiamanis_wagg_dervilis_worden_2021}. The results of a VoI/VoSHM analysis largely depend on prior knowledge related to the different available models (see \cref{sec:models}), operational conditions, environmental and load variabilities, as well as the anticipated type of damages and mitigation options that are of relevance for decision support.
A VoI/VoSHM analysis can be used as a tool to support decisions on whether or not to invest in installation, or re-configuration of an SHM system \citep{KAMARIOTIS2023109708}, to optimize its design \citep{MALINGS2016219,CANTEROCHINCHILLA2020106377, eichner2023optimal}, or to rank candidate options.
\subsection{Methods for optimizing decision strategies}
\label{subsec:methods_decision_opt}
When used in long-term monitoring settings, SHM delivers a set of data in a sequential manner at discrete points in time throughout the decision time horizon. A temporal sequence of decisions on actions needs to be optimized. This belongs to the class of stochastic Sequential Decision Problems (SDPs) \citep{Kochenderfer_2022}. The principle of maximum expected utility still applies, however, optimizing a strategy involves taking into account future actions and observations. Solution to stochastic SDPs can be cumbersome and calls for large computational efforts. 

Let us consider a simplistic decision setting, where one has to decide at each time step $t_k, \ k=1,\dots,n_T$ throughout a component's life-cycle $T$ whether to repair (R) a component or do nothing (DN), in view of continuous monitoring information (i.e., monitoring data are available at each $t_k$). The set of possible actions at time $t_k$ is $a_k=\{\text{R, DN}\}$. The sequence of actions throughout the life-cycle is $\{a_1,\dots,a_{n_T}\}$. Monitoring data obtained at each time step affect the repair decision. In turn, the repair decision affects the state of the component, and consequently also the decisions at future points in time. This simple example aims to demonstrate the complex nature of stochastic SDPs.

Numerous frameworks and algorithms are available for solving stochastic SDPs \citep{Kochenderfer_2022}. 
In the context of maintenance planning for structures and infrastructure systems, frameworks that have been employed for the solution of SDPs include heuristic decision policies \citep{Luque_2019,BISMUT2021107891}, Markov decision processes (MDPs)/partially observable Markov decision processes (POMDPs) \citep{Papakonstantinou_2014,MEMARZADEH_2016,Morato_2022,SONG2022108034}, and deep reinforcement learning (RL) \citep{Andriotis_2019,arcieri2023pomdp}. To effectively transfer these frameworks in real-world applications, it is crucial to ensure that the solutions they offer are interpretable, safe, and adhere to operational constraints \citep{ANDRIOTIS2021107551}.

\section{Increasing value creation for SHM}
\label{sec:quantifying}
To date, for many structures and infrastructure systems, the O\&M process is mainly based on an \textit{ad-hoc} usage of data. This makes it difficult to demonstrate the effects of SHM on the life-cycle costs and the performance of the systems, and to integrate it into standard operations and regulations. We thus identify two main directions in which progress is needed to enhance the SHM value generation. Firstly, since SHM provides its value by improving the decision-making process, a more explicit consideration of the way in which SHM informs decisions on O\&M of systems is required. Secondly, improved Verification and Validation (V\&V) of SHM is essential for wider usage and acceptance. 

\subsection{Integrating SHM into the decision-making process}
Shifting to a data-driven and algorithmic-driven management entails convincing owners and operators, as well as policy makers, to adopt advanced SHM technologies and trust decision algorithms. This is a challenging process, as for many systems it requires a shift from historically trusted rule-based inspection and maintenance regimes to unproven performance-based philosophy and regulations. This shift is particularly difficult for safety-critical functionalities, where failures can lead to injuries and loss of life and where prescriptive regulations often leave little room for reducing the current level of inspections and maintenance activities. 

 It is crucial to better understand and formalize the maintenance strategies that are currently in place, to facilitate a direct comparison of SHM-supported decision processes against the existing (usually empirical) approaches that the operators currently adopt. Getting the stakeholders involved in understanding and formalizing the decision-making challenges can help in better defining the utility function for decision support (e.g., by taking into account the aversion of operators to unforeseen downtime). 

Decision makers are often reluctant to adopt SHM because they assume that their current traditional O\&M strategy must be completely transformed by this adoption, and that the rationale of the resulting strategy will not be fully comprehensible to them. However, the integration of SHM data in the decision strategy can occur gradually and in a controlled process. To investigate the benefit of SHM, one can first assess the efficacy of the strategy currently adopted by the decision makers, according to the excepted utility metric (as illustrated above), and identify what changes are suggested on availability of SHM data. By simulating evolution scenarios, one can assess the effectiveness of “intermediate” strategies, which integrate the traditional assessment regimes (e.g., visual inspection) and monitoring-driven suggestions. Then, the decision maker can gradually implement some intermediate strategy and, depending on the empirical effectiveness, as evidenced in the field, select the appropriate integration level.

Provenance of predictions provided by prognostic algorithms must be presented in an "understandable manner" to practitioners. Automated cost-effective modeling processes and standardized options for instrumentation could help lower the cost concern on the operators' side. Operators also need an explicit link between SHM-based diagnostic/prognostic indicators and types of actions that are to be taken. The methods presented in \cref{subsec:methods_decision_opt} can also provide this important mapping, from data to actions. Performance is also a key issue as too many false alarms will undermine trust in the SHM system. 

VoI/VoSHM analyses, as introduced in \cref{subsec:VOI}, determine the economic benefit of deploying SHM on structures and infrastructure systems. Reliable prognosis and models are needed for accurately quantifying the VoI, yet even when models are inaccurate, VoI computation can be treated as an optimization problem, informing SHM practices and options. 
VoI/VoSHM estimates are often characterized by a large variability. An additional complication stems from the fact that cost variables of a decision problem are themselves uncertain. Stakeholders can again assist in defining preferences and costs for potential consequences of different events and actions. It is essential that VoI analyses are made transparent and convincing. There exist certain systems where an obvious value exists in monitoring for preventing failure. There, SHM can be supported even in absence of proof of VoI. Furthermore, putting a precise number on the VoI is not as important as ensuring transition from limited information and \textit{ad-hoc} decisions to knowledge and data-supported decision-making.

\subsection{Verification and validation}

Verification \& Validation (V\&V) is an essential, yet especially challenging process to establish trust in the decision support capabilities of SHM \citep{ThackerOct2004}. As of yet, V\&V of SHM is not a well-understood process, even in the more well-defined context of condition monitoring for industrial assets. It requires the definition of high-level requirements and then subsequently cascading these to lower tiers and eventually down to the most granular levels, specifying the requisites for the performance of SHM algorithms \citep{Saxena_2013}. 
The section below describes the role and interplay of SHM system requirements, system design, V\&V, and finally operations.

\textbf{System Requirements:} 
High-level system requirements comprise functional prerequisites as well as non-functional performance criteria (e.g., safety, availability) and cost requirements related to factors like damage, unscheduled maintenance, and downtime, among others.
It is essential to establish methods for testing and verifying compliance with these requirements, which often results in the creation of testability prerequisites. These testability requirements may then extend to the development of simulation models, testbeds, built-in-test modules, and supplementary testing resources to be utilized during the verification phase. 

\textbf{Detailed System Design:}
This stage necessitates failure, risk and reliability analyses
to identify performance targets for health management, which should influence the chosen SHM architecture. SHM design, when chosen to be deployed from cradle-to-grave, must adhere to constraints cascading down from the overall system design and operational requirements. These can include specifications for model fidelity and computational complexity (when considering a hybrid setting and use of a DT), sensor resolution, power requirements, sampling rates, and more.


\textbf{System Verification and Validation:}
A SHM system encompasses the implementation of hardware and software components for managing relevant sensors, signal conditioning/processing, and health management (diagnostic and prognostic) algorithms. Ideally, during a verification phase, the SHM system can be experimentally tested within 
supporting test platforms and tools required for testing. In the PHM domain, and the monitoring of industrial components and assets, it is common practice to simulate the relevant environment at the system level to ensure the proper functioning of the integrated system during various levels of operation, including injection of faults. It is obvious that in the case of structural systems such tests at actual scale are practically infeasible. This is where experimentation (in the form of scaled testing, or hybrid-simulation \citep{Gao_2013}) can form a crucial tool for V\&V. Perhaps the highest value can be gained from establishing and openly sharing data and experience gained from actual-scale monitoring benchmarks \citep{Peeters_Z24,Maes_2021}.
Given the uncertainty that is inherent in the systems on which SHM is applied (incentivizing use of a SHM system), enhanced validation of SHM can be achieved through a wide-spread application to a larger portfolio of structures or components.

\textbf{Operation and Maintenance:}
Assessment metrics should be in place to measure SHM performance over time. The employed models will typically require continual updating as both the structural and sensing system parameters change over time. Parameters and thresholds related to SHM may need fine-tuning as more information becomes available regarding the system's response to environmental and operational variabilities. Moreover, as improved technologies become available, there may be a desire to apply updates to the SHM system as well. Here, the VoI concept can serve as a potent tool for quantifying the value of choices pertaining to system modifications or upgrades.

\section{Concluding remarks}
\label{sec:concluding}
This paper underscores the potential of Structural Health Monitoring (SHM) systems to support decisions for operation and maintenance (O\&M) of structures and infrastructure systems, while also discussing multiple challenges that arise in the process of materializing SHM in these contexts. 

We first present a novel classification of SHM use cases along some principal dimensions that relate to the nature of the decision-making task. Secondly, we describe model classes that are required for enabling monitoring-informed condition-based or predictive O\&M planning. Specifically, we touch upon purely data-driven models and hybrid models (including digital twins), and we comment on the suitability of each model class in relation to the engineering system's characteristics (e.g., uniqueness, safety-criticality) and the corresponding type of the available monitoring data (e.g., sufficient experimental data to failure or not). Subsequently, we discuss the optimization of O\&M decision strategies under the availability of monitoring data and we describe relevant computational frameworks and the Bayesian decision analysis scheme, which forms the basis for Value of Information (VoI) and Value of SHM (VoSHM) analyses.

Finally, we identify two key avenues to enhance the value of SHM. The first avenue advocates the deepening of our understanding on the means and ways by which SHM impacts the decision-making process. This is a crucial process that requires the research community and the industry stakeholders to come together to address several challenges. The latter usually show reluctance to adopt SHM systems, but their input is defining as part of the SHM-informed decision support process, e.g., via formalizing the maintenance strategies that are currently in place and via determining the most relevant decision problems. The second avenue focuses on rethinking the Verification \& Validation (V\&V) process within the SHM context. Albeit a challenging and often poorly comprehended procedure, V\&V is of paramount importance towards a broader acceptance and exploitation of SHM.

We believe that interdisciplinary, collaborative efforts, similar to the one reflected in this position paper, are key to reaching a reliable synergy
between SHM and decision-making.

\begin{Backmatter}

\paragraph{Funding Statement} This work received no specific grant from any funding agency, commercial or not-for-profit sectors.

\paragraph{Competing Interests}
The authors declare no competing interests exist.

\paragraph{Data Availability Statement} Data availability is not applicable to this article as no new data were created or analyzed in this study.

\paragraph{Author Contributions}
Conceptualization: A.K, E.C., D.S., N.D., K.G, A.J.H., G.L, C.P, K.G.P., M.P., M.T., K.W.; Writing - original draft: A.K.; Writing - review and editing: E.C., D.S., N.D., K.G, A.J.H., G.L, C.P, K.G.P., M.P., M.T., K.W.; All authors approved the final submitted draft.

\bibliographystyle{apalike}
\bibliography{mybibfile}

\end{Backmatter}

\end{document}